\documentstyle[prl,aps,epsf,multicol,amssymb]{revtex}

\def\dirbib{/home/exp/e13/jwuttke/tx/bib/}
\def\dirfig{}

\begin{document}
\hyphenation{extra-po-la-tion para-meters}

\title{\boldmath Calcium Rubidium Nitrate:
       Mode-Coupling $\beta$ Scaling without Factorization}

\author{
        M.~Goldammer$^1$\cite{x59}, 
        C.~Losert$^{1}$\cite{x60},
        J.~Wuttke$^{1}$\cite{x61},
        W.~Petry$^1$,
        F.~Terki$^2$, 
        H.~Schober$^{1,3}$, 
        P.~Lunkenheimer$^4$
       }

\address{$^1$ Physik-Department E13, Technische Universit\"at M\"unchen, 
         85747 Garching, Germany}
\address{$^2$ Laboratoire de Science des Mat\'eriaux Vitreux,
          Universit\'e Montpellier II, 34095 Montpellier Cedex, France}
\address{$^3$ Institut Laue-Langevin, 38042 Grenoble Cedex 9, France}
\address{$^4$ Experimentalphysik V, 
         Universit\"at Augsburg, 86135 Augsburg, Germany}

\date{submitted to Phys.~Rev.~E: \today}

\maketitle

%%%%%%%%%%%%%%%%%%%%%%%%%%%%%%%%%%%%%%%%%%%%%%%%%%%%%%%%%%%%%%%%%%%%%%%%%%%%%%
\begin{abstract}
The fast dynamics of viscous calcium rubidium nitrate is investigated
by depolarized light scattering, neutron scattering and dielectric
loss. Fast $\beta$ relaxation evolves as in calcium potassium
nitrate. The dynamic susceptibilities can be described by the
asymptotic scaling law of mode-coupling theory with a shape parameter
$\lambda = 0.79$; 
the temperature dependence of the amplitudes
extrapolates to $T_c\simeq 378$~K. 
However, the frequencies of the minima of
the three different spectroscopies never coincide, in conflict with
the factorization prediction, 
indicating that the true asymptotic regime is unreachable. 
\end{abstract}
%%%%%%%%%%%%%%%%%%%%%%%%%%%%%%%%%%%%%%%%%%%%%%%%%%%%%%%%%%%%%%%%%%%%%%%%%%%%%%%

%\rightskip0.4\textwidth
\begin{multicols}{2}

%%%%%%%%%%%%%%%%%%%%%%%%%%%%%%%%%%%%%%%%%%%%%%%%%%%%%%%%%%%%%%%%%%%%%%%%%%%%%%%
\section{Introduction}
%%%%%%%%%%%%%%%%%%%%%%%%%%%%%%%%%%%%%%%%%%%%%%%%%%%%%%%%%%%%%%%%%%%%%%%%%%%%%%%

%%%%%%%%%%%%%%%%%%%%%%%%%%%%%%%%%%%%%%%%%%%%%%%%%%%%%%%%%%%%%%%%%%%%%%%%%%%%%%%
\subsection{Motivation}
%%%%%%%%%%%%%%%%%%%%%%%%%%%%%%%%%%%%%%%%%%%%%%%%%%%%%%%%%%%%%%%%%%%%%%%%%%%%%%%

Using three different spectroscopies, we have investigated
the fast dynamics of glass-forming calcium rubidium nitrate.
The work is motivated by unexpected differences between the
high-frequency dielectric loss of calcium rubidium 
nitrate and that of its homologue calcium potassium nitrate
\cite{LuPL97}. 

The mixed salt calcium potassium nitrate (composition
[Ca(NO$_3$)$_2$]$_{0.4}$[KNO$_3$]$_{0.6}$, abbreviation CKN) is one
of the best studied glass formers. In particular it was among the
first materials for which the relevance of mode-coupling theory (MCT)
\cite{Got91} was demonstrated
\cite{MeKF87a,MeKF87b,KnMF88,LiDC92a,CuDF93a,YaNe96}. In the 
cross-over region between microscopic dynamics and $\alpha$
relaxation, neutron \cite{KnMF88} and light \cite{LiDC92a} scattering
experiments could be described by the asymptotic scaling function 
of fast $\beta$ relaxation. 
In this regime any dynamic susceptibility is expected
to converge towards the same asymptote, which is 
determined by a single parameter $\lambda$.
For CKN, a value $\lambda=0.81$ was found \cite{LiDC92a}. 

Although the theory allows any $\lambda$ between 0.5 and 1.0,
experiments on many other liquids and on colloids, as well as
simulations and numeric solutions of MCT for model systems 
yield almost always values between 0.7 and 0.8.
The major exception from this has been reported for calcium rubidium 
nitrate (composition [Ca(NO$_3$)$_2$]$_{0.4}$[RbNO$_3$]$_{0.6}$,
abbreviation CRN):
Broad-band dielectric measurements suggested $\lambda=0.91$ \cite{LuPL97}.
This is unexpected because of the close similarity between CRN and
CKN \cite{ang66,PiLN97}, and it is interesting also because
for $\lambda \rightarrow 1$ one expects logarithmic corrections
to the scaling laws of MCT \cite{GoSj89b}.
In order to investigate this anomaly in more detail, we have performed
light and neutron scattering experiments and reanalyzed dielectric
data.

%%%%%%%%%%%%%%%%%%%%%%%%%%%%%%%%%%%%%%%%%%%%%%%%%%%%%%%%%%%%%%%%%%%%%%%%%%%%%%%
\subsection{\boldmath Fast Dynamics and $\beta$ Relaxation}\label{Stheory}
%%%%%%%%%%%%%%%%%%%%%%%%%%%%%%%%%%%%%%%%%%%%%%%%%%%%%%%%%%%%%%%%%%%%%%%%%%%%%%%

To explain the focus of the present study, let us anticipate some
experimental data. Figure 1 shows the dynamic susceptibility of CRN as
measured by depolarized light scattering. The temperature-independent
band on the high-frequency side is attributed to the
microscopic dynamics: Vibration, rotation, 
librations. Above 3 THz the susceptibilities coincide for all
temperatures, as expected for purely harmonic excitations. At lower
frequencies, on the other hand, the dynamics is clearly
anharmonic. In the low-frequency, high-temperature limit the curves
bend over towards a maximum which itself remains outside the experimental
frequency window. This maximum is due to structural $\alpha$ relaxation. 
It is strongly temperature dependent;
its evolution towards lower temperatures and lower frequencies can be
studied by many other, slower spectroscopic techniques.

In the following we concentrate on the intermediate frequency range
 around the susceptibility minimum. In this regime, called {\em fast
 $\beta$ relaxation}, one finds nontrivial contributions to the
 dynamic susceptibility, which can
 {\em not} be explained by a simple superposition
of $\alpha$ relaxation and microscopic excitations. 
The existence of this regime 
has confirmed a key result of mode-coupling theory. 

More specifically:
for temperatures little above a critical temperature, $T\gtrsim T_c$, 
and for frequencies around the susceptibility minimum, $\nu\sim\nu_\sigma$,
any susceptibility is predicted to converge towards the same asymptote 
\begin{equation}\label{Echi}
 \chi''(\nu) \simeq \chi_\sigma~g_\lambda(\nu/\nu_\sigma).
\end{equation}
This prediction is often described as a {\em factorization}: For
instance, wave-number dependent neutron scattering data
$\chi''(q,\nu)$ factorize into an amplitude, which depends only on
$q$, and a spectral distribution, which depends only on the
frequency~$\nu$.
The amplitudes
\begin{equation}\label{Echisig}
 \chi_\sigma \propto |\sigma|^{1/2}
\end{equation}
and the characteristic frequency of $\beta$ relaxation
\begin{equation}\label{Enusig}
 \nu_\sigma \propto |\sigma|^{1/2a}
\end{equation}
are predicted to depend for all spectroscopies in the same way
on the reduced temperature 
\begin{equation}\label{Esigma}
 \sigma = { T_c - T \over T_c }.
\end{equation}
The two wings of the susceptibility minimum are described by power
laws
\begin{equation}\label{Epower}
 g_\lambda({\tilde{\nu}}) \propto 
  \left\{ 
     { \tilde{\nu}^{-b} \atop \tilde{\nu}^{a\hphantom{-}} } \quad
     { \mbox{for} \atop \mbox{for}  } \quad
     { \tilde{\nu}\ll 1, \atop  \tilde{\nu}\gg 1. }
  \right.
\end{equation}
The fractal exponents $a$ and $b$ depend on the parameter
$\lambda$ which alone determines the full shape \cite{Got90} 
of $g_\lambda(\tilde{\nu})$. 

The predictions (\ref{Echi})--(\ref{Enusig}) have been confirmed over
a wide temperature range for the model liquid CKN. In the present
study we shall analyze how far the same description applies also to
the fast dynamics of CRN.

%%%%%%%%%%%%%%%%%%%%%%%%%%%%%%%%%%%%%%%%%%%%%%%%%%%%%%%%%%%%%%%%%%%%%%%%%%%%%%
%%%                                                                        %%%
%%%     II. Experiments and raw data treatment                             %%%
%%%                                                                        %%%
%%%%%%%%%%%%%%%%%%%%%%%%%%%%%%%%%%%%%%%%%%%%%%%%%%%%%%%%%%%%%%%%%%%%%%%%%%%%%%
\section{Experiments and raw data treatment}

%%%%%%%%%%%%%%%%%%%%%%%%%%%%%%%%%%%%%%%%%%%%%%%%%%%%%%%%%%%%%%%%%%%%%%%%%%%%%%
\subsection{Samples}
%%%%%%%%%%%%%%%%%%%%%%%%%%%%%%%%%%%%%%%%%%%%%%%%%%%%%%%%%%%%%%%%%%%%%%%%%%%%%%

The mixture of calcium nitrate and rubidium nitrate,
[Ca(NO$_3$)$_2$]$_{0.4}$[RbNO$_3$]$_{0.6}$, was prepared as in
Ref.~\cite{LuPL97}.  
For the light-scattering measurements the sample material was transferred
under helium atmosphere into a glass cuvette, which was sealed, and
mounted in a cryofurnace that can reach temperatures up to
470$~$K. For the neutron-scattering measurement at the Institut
Laue-Langevin (ILL) the material was filled into an aluminum container
\cite{Wut99,Wut99E} so that the sample forms a hollow cylinder of
thickness  1$~$mm, outer radius 29$~$mm, and height  50$~$mm; the
aluminum itself was 0.1$~$mm thick. The cell was mounted in  an ILL
cryo loop; a thermocouple in direct contact with the sample was used
to determine the temperature.  

The samples showed little tendency to crystallize, except when kept
for several hours between 400 and 410$~$K. The glass state can be
reached easily by supercooling with moderate speed ($\lesssim
5~$K/min).
The glass transition temperature $T_g = 333~$K \cite{PiLN97} of CRN
is exactly the same as of CKN. The melting point of crystalline CRN
has not been measured, but we suppose that it is similar to $T_m\simeq
438~$K of CKN.

%%%%%%%%%%%%%%%%%%%%%%%%%%%%%%%%%%%%%%%%%%%%%%%%%%%%%%%%%%%%%%%%%%%%%%%%%%%%%%
\subsection{Depolarized light scattering}
%%%%%%%%%%%%%%%%%%%%%%%%%%%%%%%%%%%%%%%%%%%%%%%%%%%%%%%%%%%%%%%%%%%%%%%%%%%%%%

Light scattering was measured on a six-pass tandem
interferometer and a two-pass grating monochromator.
A near back-scattering (173$^\circ$) geometry with crossed polarizers
(HV) was used to minimize scattering from acoustic modes. Stray light
was negligible compared to the dark counts of the avalanche
photodiodes. The dark count rate of about 2.5$~$sec$^{-1}$ 
was in turn more than 10 times weaker than the weakest
scattering signals; nevertheless we subtracted it from all measured
spectra. 

For the high-frequency regime from 100$~$GHz to 5$~$THz we used
a Jobin-Yvon U1000 double monochromator. The
entrance and exit slits were set to 50$~\mu$m and the slits between the
two monochromators to 100$~\mu$m, resulting in a full width at
half maximum of 10 GHz and a straylight suppression of better than
$10^{-5}$ at 100 GHz. To maintain this resolution over hours, days,
and months the temperature of the instrument has to be stabilized
carefully ($\pm 0.05~$K); therefore the monochromator is placed in an
insulating box ($<0.5~$W/(m$^2\cdot$K)) and thermalized by forced
convection: The air temperature inside the box is regulated by an
electrical heater placed in front of a fan, acting against large
water-cooled copper sheets. 

For the low frequency regime from 0.7 to 180$~$GHz we used a
Sandercock-Fabry-Perot six--pass tandem interferometer, which we modified in
several details as described in Ref.~\cite{WuOG00a}, in order to allow
for stable operation and high contrast. 
The instrument was used in series with an interference filter of
either 150 or 1000$~$GHz bandwidth that suppresses higher-order
transmission leaks of the tandem interferometer
\cite{SuWN98,GaSP99,BaLS99} below 3\% or better. The filters are
placed in a specially insulated housing with active temperature
stabilization. To account for any drift, the instrument function is
redetermined periodically by automatic white-light scans.

In the present study we used three different free spectral ranges (188,
50, and 16.7$~$GHz), corresponding to mirror spacings of 0.8, 3, and
9$~$mm. Measured spectra were divided by the white-light
transmission. Then they were
converted from intensity to susceptibility 
\begin{equation}\label{Echils}
\chi''_{\rm ls}(\nu)=I(\nu)/n(\nu)
\end{equation}
 with the Bose factor $n(\nu) =
\left[\exp(h\nu/k_{\rm B}T)-1\right]^{-1}$;
this representation is more sensitive
to experimental problems. Whenever Stokes and anti-Stokes data did
not fall onto each other the measurement was repeated. Finally scans
taken at different spectral ranges were joined after adjusting the
intensity scales by a least-square match of the overlapping data
points. Additional scans at other mirror spacings confirmed the
accuracy of the composite broad-band susceptibilities.

The overall intensity scale was taken from an unperturbed temperature
cycle on the interferometer with a free spectral range of 188~GHz, to
which all other measurements were matched. The
accuracy of this procedure was confirmed in the THz range, where
harmonic vibrations are expected to yield a temperature-independent
susceptibility: In fact our $\chi''_{\rm ls}(\nu)$ coincide within 4\%.

The temperature independence of $\chi''_{\rm ls}(\nu)$ in the
vibrational range was used to set the intensities of the 400 and 410~K
spectra which had to be measured separately because of the tendency to
crystallization.

%%%%%%%%%%%%%%%%%%%%%%%%%%%%%%%%%%%%%%%%%%%%%%%%%%%%%%%%%%%%%%%%%%%%%%%%%%%%%%
\subsection{Neutron scattering}
%%%%%%%%%%%%%%%%%%%%%%%%%%%%%%%%%%%%%%%%%%%%%%%%%%%%%%%%%%%%%%%%%%%%%%%%%%%%%%

Figure~\ref{ns01}a shows the static structure factor~$S(q)$,
measured by neutron diffraction at about room temperature,
of CRN (a rapid measurement on G41 at the LLB)
and of CKN (from the detailed study~\cite{KaCC95b}).
Except for the amplitude,
no adjustments have been made.
Over most of the $q$~range,
the $S(q)$ of both materials agree exceptionally well;
the deviations below 1.2~\AA$^{-1}$ are most probably due to
multiple scattering and other artifacts
which are expected to affect mostly the low-$q$ region.
Thus we can detect no difference between the structures of CRN and CKN,
although the cation diameters 
of Rb$^{+}$ (2.94~\AA) and K$^{+}$ (2.66~\AA) differ substantially.

Inelastic neutron scattering was measured on the time-of-flight spectrometer
IN~5 at the ILL. This is a multi-chopper instrument which allows for
free choice of the incident neutron wavelength $\lambda_i$. With
increasing $\lambda_i$, the width of the instrumental resolution
improves as $\lambda_i^{-3}$, on the expense of decreasing flux and
decreasing wave numbers. As a compromise, we choose
$\lambda_i=8.0~\rm\AA$. 
This yields a resolution (FWHM) of 8~GHz.
Depending on temperature, 
we cut the inelastic data at typically 20~GHz,
where the signal-to-noise ratio is of the order of~$10^3$.

Our choice of $\lambda_i$ restricts the wave number range for elastic
scattering to $q\lesssim1.3~\rm\AA^{-1}$. Thus we do not
reach the maximum of the structure factor at 1.9$~\rm\AA^{-1}$.
Therefore we get rather low scattering intensities; 
fortunately, the static structure factor maximum is also inaccessible
for almost all multiple-scattering events \cite{RuML00}. 
 Figure~\ref{ns01}b shows the dynamic window of our measurement in the
$q,\nu$ plane. The accessible $q(\nu,2\vartheta)$ are shown for every
second detector angle $2\vartheta$. The gaps in the plot are due to
struts inside the flight chamber of the instrument. 

The measured $S(\nu,2\vartheta)$ are interpolated to constant $q$. Only wave
numbers between 0.5 and 1.3~$\rm\AA^{-1}$ are used in the analysis.
As a last step the $S(q,\nu)$ are converted into a susceptibility
%\begin{equation}\label{Esushi}
 $\chi''_q(\nu) = S(q,\nu)/n(\nu)$.
%\end{equation}

%%%%%%%%%%%%%%%%%%%%%%%%%%%%%%%%%%%%%%%%%%%%%%%%%%%%%%%%%%%%%%%%%%%%%%%%%%%%%%
%%%                                                                        %%%
%%%             III. Results And Analysis                                  %%%
%%%                                                                        %%%
%%%%%%%%%%%%%%%%%%%%%%%%%%%%%%%%%%%%%%%%%%%%%%%%%%%%%%%%%%%%%%%%%%%%%%%%%%%%%%

\section{Results And Analysis}\label{Sresults}

%%%%%%%%%%%%%%%%%%%%%%%%%%%%%%%%%%%%%%%%%%%%%%%%%%%%%%%%%%%%%%%%%%%%%%%%%%%%%%
\subsection{Light-scattering results}
%%%%%%%%%%%%%%%%%%%%%%%%%%%%%%%%%%%%%%%%%%%%%%%%%%%%%%%%%%%%%%%%%%%%%%%%%%%%%%

\begin{figure}
\epsfxsize=78mm\centerline{\epsffile{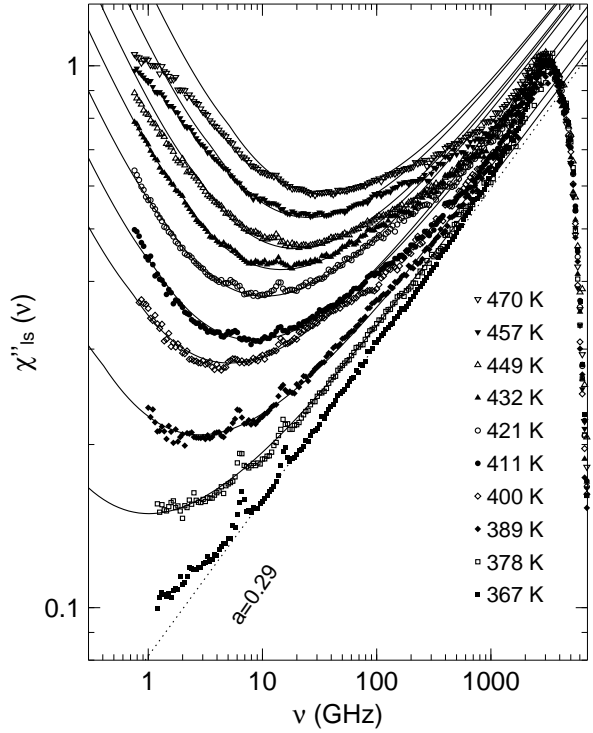}}
\bigskip
 \caption{
  Susceptibilities of CRN as measured by light scattering
  between 367~K (bottom) and 470~K (top). 
  The small peaks at about 7 and 15 ~GHz are due to
  residual TA and LA Brillouin scattering. For $\nu\gtrsim 3~$THz all
  data fall together, as expected for harmonic vibrations.
  In the low-frequency, high-temperature limit the curves bend towards
  the $\alpha$-relaxation peak. 
  The intermediate regime of fast $\beta$ relaxation can be described
  by the asymptotic scaling function of mode-coupling theory. 
  The solid curves show fits~(\ref{Echi}) with a common shape
  parameter $\lambda=0.79$.
  The dotted line shows a power-law fit to the lowest temperature from 
  $10$ to $1000~\rm{GHz}$. The resulting exponent of $a=0.29$ 
  corresponds to $\lambda=0.78$.}
  \label{ls02}
\end{figure}

The light scattering data have already been presented in
Sect.~\ref{Stheory} (Fig.~\ref{ls02}). We now concentrate on the 
intermediate regime of fast $\beta$ relaxation. At low temperatures
the high-frequency wing of the susceptibility minimum follows a power
law $\chi''_{\rm ls}(\nu) \propto \nu^a$. At 367$~$K this power law
extends over two decades from 10 to 1000$~$GHz with $a=0.29$. Within
MCT this exponent corresponds to a shape parameter $\lambda =
0.78$. This has to be compared to the dielectric measurements \cite{LuPL97}
where a similar power-law behavior has been observed over more than three
decades with $a=0.2$ for 361$~$K, leading to the exceptional value
$\lambda = 0.91$. 

Looking for the power-law asymptotes (\ref{Epower}) is obviously not
the best way of testing the applicability of MCT. The scaling
prediction (\ref{Echi}) is expected to hold best in the intermediate
regime around $\nu \sim \nu_\sigma$. Therefore we use the full
scaling function $\chi_\sigma g_\lambda(\nu/\nu_\sigma)$ to fit the
experimental data around the minimum. Figure 2 shows such fits for two
temperatures and with three different values of $\lambda$. From this
comparison we obtain $\lambda = 0.79$ with an accuracy better than
$\pm 0.01$. Figure~\ref{ls02} contains fits with fixed $\lambda=0.79$
for all temperatures.

From these fits we extract the amplitude $\chi_\sigma$ and the
characteristic frequency $\nu_\sigma$ as functions of temperature. For
constant $\lambda$, the parameters $\chi_\sigma$ and $\nu_\sigma$ are
proportional to the height and the position of the susceptibility
minimum. In order to test the predictions (\ref{Echisig}) and
(\ref{Enusig}), Figure~\ref{ls04} shows them as $\chi_\sigma^2$ and
$\nu_\sigma^{2a}$. Linear fits to the lowest five points 
confirm the predictions (\ref{Echisig}),(\ref{Enusig}) over an
interval of 50$~$K and extrapolate consistently to a value for $T_c$
between about 365$~$K and 370$~$K. 

%%%%%%%%%%%%%%%%%%%%%%%%%%%%%%%%%%%%%%%%%%%%%%%%%%%%%%%%%%%%%%%%%%%%%%%%%%%%%%
\subsection{Neutron scattering results}
%%%%%%%%%%%%%%%%%%%%%%%%%%%%%%%%%%%%%%%%%%%%%%%%%%%%%%%%%%%%%%%%%%%%%%%%%%%%%%
 Compared to depolarized light scattering, neutron scattering features
 as an additional coordinate the wave number $q$. This allows for a
 direct test of the factorization property (\ref{Echi}): In the
 asymptotic regime of fast $\beta$ relaxation the susceptibility
 $\chi''_{\rm ns}(q,\nu)$. Such a factorization is also expected for
 one-phonon scattering from a harmonic system.

 As described elsewhere \cite{WuOG00a,WuKB93} the $h_q$ are determined
 from a least-square match of neighboring $q$ cuts. Figure~\ref{ns02}
 shows the rescaled $\chi''_{\rm ns}(q,\nu)/h_q$; the inset shows the
 $h_q$. As in other cases the $h_q$ do not go with $q^2$ which can be
 explained by an almost $q$-independent multiple-scattering background
 \cite{Wut00c}.
 
 The factorization holds around the
 $\beta$ minimum as well as for the vibrational band; only for the highest
 temperature the $q$-dependent $\alpha$ peak comes into the
 experimental window. This allows us to collapse all $q$ cuts into an
 average susceptibility
 \begin{equation}\label{Echins}
 \chi''_{\rm ns}(\nu) =  \left< \chi''(q,\nu) / h_q \right>_q
 \end{equation}
 with much improved statistics. Results are shown in
 Figure~\ref{ns03}.

 Fits with the mode-coupling asymptote $\chi_\sigma
 g_\lambda(\nu/\nu_\sigma)$ 
 allow for any value of $\lambda$ between 0.7 and 0.8, but
 definitely not for $\lambda = 0.91$. Therefore we impose the
 light-scattering result $\lambda = 0.79$. For most temperatures, the
 fits describe the susceptibilities over one decade or more. 
 
 Towards low frequencies, the fit range is restricted by the
 instrumental resolution, except for the highest temperatures where
 $\alpha$ relaxation is resolved (Fig.~\ref{ns02}). On the
 high-frequency side, the fit range extends up to about 200$~$GHz.
 At higher frequencies neutron-scattering data rise {\em above} the fit, 
 whereas light-scattering data fell 
 {\em below} the theoretical curves. 
This gives an upper limit for the frequency range
of the asymptotic regime of fast $\beta$ relaxation.

 A rectified plot  of the $\beta$-relaxation parameters is shown in
 Figure~\ref{ns04}. The amplitude $\chi_\sigma$ follows the power law
 (\ref{Echisig}) over 140$~$K, a much wider temperature interval
 than in light scattering. The straight line fitted to
 $\chi_\sigma^2$ extrapolates to $T_c \simeq 380~$K. 
 On the other hand, the frequency $\nu_\sigma$ does not obey
 (\ref{Enusig}). At high temperatures, the $\nu_\sigma^{2a}$ seem to
 lie on a line which extrapolates to an unphysical $T_c$ far
 below $T_g$. For lower temperatures, the data possibly bend over
 towards the true asymptote which however is not reached in our
 experiment. At this  point we should note that the $\nu_\sigma$ are
 much more sensitive to  noise and to remnants of the instrumental
 resolution than the $\chi_\sigma$. 

%%%%%%%%%%%%%%%%%%%%%%%%%%%%%%%%%%%%%%%%%%%%%%%%%%%%%%%%%%%%%%%%%%%%%%%%%%%%%%
\subsection{Reanalyzing dielectric data}
%%%%%%%%%%%%%%%%%%%%%%%%%%%%%%%%%%%%%%%%%%%%%%%%%%%%%%%%%%%%%%%%%%%%%%%%%%%%%%

 Dielectric loss has been measured in CRN over more than 11 decades in
 frequency, from 1$~$mHz to 380$~$GHz \cite{LuPL97}. Here we
 concentrate on the fast $\beta$ relaxation.
 The most remarkable feature of this regime
 is the extremely slow increase of $\epsilon''(\nu)$ on
 the high-frequency side of the minimum.
 For three decades in frequency ($40\mbox{ MHz} - 40\mbox{ GHz}$)
 $\epsilon''(\nu)$ follows a power law with an exponent $a=0.2$. 
 As anticipated above, this implies 
 $\lambda=0.91$. Using this value, the dielectric data could be fitted
 with the scaling function $\epsilon_\sigma g_\lambda
 (\nu/\nu_\sigma)$ over a wide range, extending from about the minimum
 up to the highest measured frequencies.

 As we have seen above, $\lambda=0.91$ is not compatible with the
 light and neutron scattering results. Furthermore we have seen that
 the asymptotic regime does not extend above some 100~GHz. Therefore we
 now reanalyze the dielectric data with an imposed value
 $\lambda=0.79$, concentrating on lower frequencies. The resulting
 fits are shown in Figure \ref{dk01}. As expected, the fits do not
 match the high-frequency wings (the data fall below the fit function,
 as in light scattering); on the other hand, the fits
 now cover much of the $\nu<\nu_\sigma$ wing.

 The two fit parameters are shown in Figure~\ref{dk02}.
 The $\epsilon_\sigma^2$ suggest $T_c\simeq377~$K.
 Within their experimental uncertainty,
 the $\nu_\sigma^{2a}$ seem compatible with such an extrapolation;
 their determination suffers however from the hitherto unavoidable
 frequency gaps in the dielectric broadband measurements.

%%%%%%%%%%%%%%%%%%%%%%%%%%%%%%%%%%%%%%%%%%%%%%%%%%%%%%%%%%%%%%%%%%%%%%%%%%%%
%%%                                                                      %%%
%%%             IV. Direct comparison of results                         %%%
%%%                                                                      %%%
%%%%%%%%%%%%%%%%%%%%%%%%%%%%%%%%%%%%%%%%%%%%%%%%%%%%%%%%%%%%%%%%%%%%%%%%%%%%

\section{Comparison of the Three Spectroscopies}

For each of the three spectroscopic techniques employed in this study,
we found a fast $\beta$ relaxation. For each of the three data sets,
we verified the asymptotic validity of the scaling function
(\ref{Echi}), and we found the temperature-dependent parameters  at
least in partial accord with the power laws (\ref{Echisig})
and~(\ref{Enusig}). 

The next question is whether the fits to the individual data sets are
consistent with each other. From the factorization property of fast
$\beta$ relaxation we expect that all susceptibilities converge
towards the {\em same} frequency and temperature
dependence. Anticipating this prediction, we have already imposed 
{\em one} value $\lambda=0.79$ to the analysis of all three data
sets. Our fits confirm that all data can indeed be described by the
same scaling function $\chi_\sigma g_\lambda(\nu/\nu_\sigma)$.

Additionally we expect a consistent temperature dependence of all
amplitude $\chi_\sigma$ and frequencies $\nu_\sigma$. A fortiori,
power-law fits to these parameters must extrapolate toward a unique
value of $T_c$. In 
Sect.~\ref{Sresults} linear fits to the $\chi_\sigma^2$ gave 
\begin{quote}
\begin{tabular}{l@{~~}l@{~~}l}
\tableline
\tableline
$T_c$ &from method &range\\
\tableline
  $\simeq 367$~K &light scattering   &378--412~K\\
  $\simeq 378$~K &neutron scattering &392--530~K\\
  $\simeq 377$~K &dielectric loss    &381--420~K\\
\tableline
\tableline
\end{tabular}
\end{quote}
The spread of these $T_c$'s is definitely larger than the
uncertainty of the experimental temperature scale. In order to reach a
consistent interpretation of the three data sets, we replot in Figure
\ref{vg02} all MCT parameters on a common temperature scale. The
amplitudes are scaled by an arbitrary factor. In this
representation all $\chi_\sigma^2$ above about 390~K appear
compatible with a common power law (\ref{Echisig}).
This suggests a reinterpretation of the light-scattering
data. Shifting the fit range to higher temperatures we find indeed 
\begin{quote}
\begin{tabular}{l@{~~}l@{~~}l}
\tableline
\tableline
$T_c$ &from method &range\\
\tableline
  $\simeq 379$~K &light scattering   &389--470~K\\
\tableline
\tableline
\end{tabular}
\end{quote}
as shown by the dashed line in Figure~\ref{ls04}. Thus the amplitudes
$\chi_\sigma$ can be given a consistent MCT interpretation with a
common $T_c \simeq 378\pm2~$K. 

The same is not true for the
frequencies $\nu_\sigma$. In light scattering we had found a
consistent asymptotic temperature dependence of~$\nu_\sigma$
and~$\chi_\sigma$. This accord is however lost after shifting the fit 
range to higher temperatures: A free fit to the $\nu_\sigma$ does not
help to obtain a $T_c$ above 370~K. In neutron scattering the $\nu_\sigma$
do not reach the power-law regime at all. Only in dielectric loss the
$\nu_\sigma$ are possibly compatible with $T_c\simeq378~$K.

Furthermore, within the asymptotic regime the $\nu_\sigma$ are
expected to agree in absolute value. The comparison in
Figure~\ref{vg02} shows that this is not the case for any 
temperature. This violation of the MCT factorization prediction is
also confirmed by the direct comparison of measured susceptibilities
in Figure~\ref{vg01}: The positions of the minima differ by up to a
factor 10. While each data set for itself seemed to be in good accord
with the scaling predictions of MCT it now turns out that most if not
all data are outside the true asymptotic regime.

%%%%%%%%%%%%%%%%%%%%%%%%%%%%%%%%%%%%%%%%%%%%%%%%%%%%%%%%%%%%%%%%%%%%%%%%%%%%
%%%                                                                      %%%
%%%                 Discussion                                           %%%
%%%                                                                      %%%
%%%%%%%%%%%%%%%%%%%%%%%%%%%%%%%%%%%%%%%%%%%%%%%%%%%%%%%%%%%%%%%%%%%%%%%%%%%%

\section{Comparison with Other Glass Formers}

%%%%%%%%%%%%%%%%%%%%%%%%%%%%%%%%%%%%%%%%%%%%%%%%%%%%%%%%%%%%%%%%%%%%%%%%%%%%
\subsection{CRN and CKN}
%%%%%%%%%%%%%%%%%%%%%%%%%%%%%%%%%%%%%%%%%%%%%%%%%%%%%%%%%%%%%%%%%%%%%%%%%%%%

We undertook this work with the intention of comparing
CRN to the well-studied model liquid CKN.
In planning and performing the scattering experiments
we took full advantage of the experience gained in previous
investigations, and we intentionally concentrated on the temperature
and frequency window of fast $\beta$ relaxation.
Therefore it is not surprising that by now
our CRN data are more accurate and more complete than what has been
published many years ago on CKN.

CKN was the material in which Cummins and coworkers first
discovered the self-similarity of depolarized light-scattering
spectra \cite{TaLC91a}. Subsequent broad-band measurements 
were successfully described by the scaling laws of MCT,
leading to $\lambda=0.81\pm0.05$ and $T_c=378\pm5~$K
\cite{LiDC92a}.
Later the light-scattering susceptibilities were also fitted across
$T_c$ with extended MCT \cite{CuDF93a}.
Unfortunately, these studies, as any other at that time,
had been undertaken with an unsufficient band pass
in the tandem interferometer \cite{SuWN98,GaSP99,BaLS99}.
Higher-order leaks cause distortions of the spectral line shapes
which are most harmful at low temperatures.
Above $T_c$, all qualitative observations will remain valid,
but as we have shown in the case of propylene carbonate \cite{WuOG00a}
the parameter $\lambda$ might change by as much as 0.06.

CKN was also the material in which neutron scattering experiments
by Mezei and coworkers first showed the relevance of MCT.
Elastic scans gave the first evidence for the onset of fast $\beta$
relaxation on approaching $T_c$, estimated at about 368~K
\cite{MeKF87b}. Later Mezei emphasized the uncertainty of this
determination \cite{Mez91a}.
Combined back-scattering and time-of-flight measurements revealed
the cross-over between the asymptotic power laws $S(q,\nu)\propto
\nu^{-1-b}$ and $\nu^{-1+a}$ \cite{KnMF88};
a free fit of $a$ gave $\lambda\simeq0.80$ whereas a consistent set of
$a$ and $b$ suggested $\lambda\simeq0.89$.
More recent neutron scattering experiments concentrated on
$\alpha$ relaxation \cite{KaMe95a}, on the static structure factor 
\cite{KaCC95b,KaCC96}, and on the microscopic dynamics above 100~GHz
\cite{RuML00,KaCC96,MeRu99}; 
a state-of-the-art determination of $\lambda$ and $T_c$ for CKN
is presently missing.

In contrast, dielectric loss in CKN has been measured recently \cite{LuPL97}
and with the same accuracy as in CRN.
MCT fits to the CKN data gave $\lambda\simeq0.76$ and $T_c\simeq375~$K.

In this situation,
it would be worthwhile to remeasure the dynamic susceptibility of CKN 
at selected temperatures above $T_c$ by light and neutron
scattering.
Such measurements would allow to determine more precise values of
$\lambda$
and $T_c$, 
to cross-check the $\beta$ relaxation parameters obtained by 
 different spectroscopies,
and then to compare in more
detail the overall behavior of CKN to that of CRN.

On the basis of the available data
we can conclude that as soon as we restrict our analysis to frequencies
below about 100~GHz
there is no significant difference in the fast dynamics of CKN and CRN.
The uncertainty in the determination of $\lambda$, especially for CKN,
is presently much larger than any difference between CKN and CRN.
Since both materials have the same calorimetric glass transition
temperature,
it is not unreasonable to compare also the $T_c$'s on absolute scale;
the value 378~K for CRN agrees perfectly well
 with the best available estimates for CKN.

%%%%%%%%%%%%%%%%%%%%%%%%%%%%%%%%%%%%%%%%%%%%%%%%%%%%%%%%%%%%%%%%%%%%%%%%%%%%
\subsection{CRN and Organic Glass Formers}
%%%%%%%%%%%%%%%%%%%%%%%%%%%%%%%%%%%%%%%%%%%%%%%%%%%%%%%%%%%%%%%%%%%%%%%%%%%%

Similar studies of fast $\beta$ relaxation have already been undertaken
in a number of organic glass formers.
Light and neutron scattering around the
susceptibility minimum have been compared in glycerol \cite{WuHL94},
salol \cite{ToPD96}, toluene \cite{WuSH98}, 
and trimethylheptane \cite{ShTB00}.
In propylene carbonate \cite{WuOG00a}
the scattering experiments have also been compared to dielectric and
time-dependent optical measurements. 

Most of these studies show the same trend as the present CRN data:
Individual data sets seem in good accord with the scaling predictions of
MCT,
but the positions of the susceptibility minima do not coincide.
The major exception is provided by glycerol
where the factorization property seems to hold although the individual
data sets do not reach the MCT asymptote \cite{WuHL94}. 

In CRN and in toluene, the susceptibility minimum lies at lower frequencies
for light scattering than for neutron scattering,
whereas in salol, trimethylheptane and propylene carbonate 
the opposite is observed.
On this basis it is presently impossible
to give any microscopic explanation \cite{ShTB00}.

We do understand why fast $\beta$ relaxation appears in the dielectric
loss data only within a rather small temperature range (up to about
40~K above $T_c$, whereas neutron scattering data show a $\beta$
minimum up to at least $T_c + 150$~K): As explained
at length for the case of propylene-carbonate \cite{WuOG00a} this is
an immediate consequence of the scaling behavior of $\alpha$
relaxation. Since the $\alpha$ peak (relative to the susceptibility in
the phonon range) is much stronger in dielectric loss than in the
scattering experiments, it covers the $\beta$ minimum at relatively
low temperatures.

%%%%%%%%%%%%%%%%%%%%%%%%%%%%%%%%%%%%%%%%%%%%%%%%%%%%%%%%%%%%%%%%%%%%%%%%%%%%%%%
\section{Conclusion}
%%%%%%%%%%%%%%%%%%%%%%%%%%%%%%%%%%%%%%%%%%%%%%%%%%%%%%%%%%%%%%%%%%%%%%%%%%%%%%%

With each material we investigate it becomes clearer that fast $\beta$
relaxation is a constitutive property of glass-forming liquids. If we
want to understand the macroscopic behavior of viscous liquids, we
will have to understand the full evolution of fast dynamics from
phonons towards $\alpha$ relaxation, passing inevitably through the
$\beta$ relaxation regime.

Studies of simple models \cite{x56} suggest that the mode-coupling ansatz is
in principle capable of providing an almost quantitatively correct
description of fast dynamics. In the last years considerable progress
has been made in extending MCT to molecular systems
\cite{FaLS00,GoSV00b,ThSL00}. A MCT of CKN 
or CRN seems not out of reach. However, until such a theory is
developed we can compare experimental results only to asymptotic
scaling laws or to numeric solutions of schematic mode-coupling models.

The present study shows once more that asymptotic fits are potentially
misleading. They provide a valuable parameterization of $\beta$
relaxation and they help to uncover the universal behavior of
different materials; but even good accord with the scaling predictions
(\ref{Echi})~--~(\ref{Enusig}) does not guarantee that the true
asymptotic regime is reached: any additional measurement can disprove
the factorization. 

This reservation does not contradict the relevance of MCT: Even for
models which obey MCT by construction it has been shown
\cite{FrFG97b,FuGM98} that the 
asymptotic $\beta$ regime is only reached at very low frequencies and
for temperatures very close to $T_c$ (where the dynamics of
molecular systems is dominated by hopping processes that are not
contained in the usual MCT). Analytic expansions beyond the first-order
scaling laws \cite{FrFG97b} make it plausible that in the preasymptotic regime
the susceptibilities are severely distorted while the amplitudes
$\chi_\sigma$ might still evolve in good accord with the power
law~(\ref{Echisig}).

In propylene carbonate it has been explicitely shown \cite{GoVo00} that the
different experimental susceptibilities can be described by
simultaneous fits with a few-parameter mode-coupling model. We have no
doubt that similar fits would also work in CRN. Considering however
that we found no qualitative difference between CRN and CKN, we
suggest that any additional experimental and theoretical effort be
invested in the generally recognized model system CKN.

%%%%%%%%%%%%%%%%%%%%%%%%%%%%%%%%%%%%%%%%%%%%%%%%%%%%%%%%%%%%%%%%%%%%%%%%%%%%%%%
\section*{Acknowledgements}
%%%%%%%%%%%%%%%%%%%%%%%%%%%%%%%%%%%%%%%%%%%%%%%%%%%%%%%%%%%%%%%%%%%%%%%%%%%%%%%

We thank A. Maiazza (TU Darmstadt) for preparing the sample material
used in all three spectroscopies, 
and G.~Andr\'e\ (LLB Saclay) for measuring the structure factor of CRN.
We acknowledge financial aid by
the Bundes\-mini\-sterium f\"ur Bildung, Wissenschaft, 
Forschung und Technologie through Verbundprojekte 03{\sc pe}4{\sc tum}9 and 
03{\sc lo}5{\sc au\footnotesize 2}8
and through contract no.\ 13{\sc n}6917,
by the Deutsche Forschungsgemeinschaft under grant no.\ {\sc Lo}264/8--1.

%%%%%%%%%%%%%%%%%%%%%%%%%%%%%%%%%%%%%%%%%%%%%%%%%%%%%%%%%%%%%%%%%%%%%%%%%%%%

\def\BIBsubm{submitted}
\def\BIBinpr{in press}
\bibliographystyle{\dirbib switch}
\bibliography{\dirbib jw1}
%here-the-bbl-file:

%%%%%%%%%%%%%%%%%%%%%%%%%%%%%%%%%%%%%%%%%%%%%%%%%%%%%%%%%%%%%%%%%%%%%%%%%%%%%

%\end{multicols}
\newpage
\narrowtext

%%%%%%%%%%%%%%%%%%%%%%%%%%%%%%%%%%%%%%%%%%%%%%%%%%%%%%%%%%%%%%%%%%%%%%%%%%%%%

\begin{figure}
\epsfxsize=78mm\centerline{\epsffile{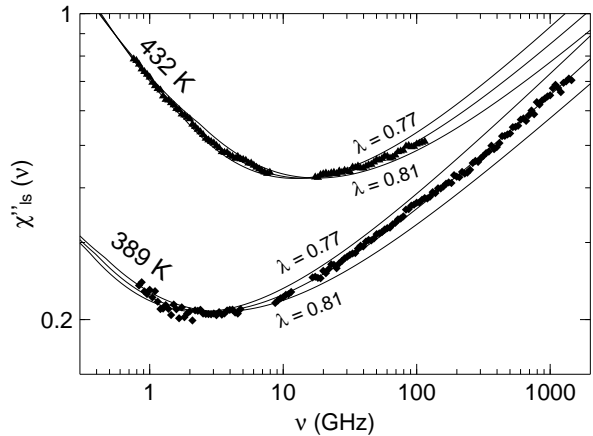}}
\bigskip
 \caption{
  A subset of the light scattering data, and mode-coupling fits with
  three different values of the shape parameter $\lambda$ (0.77, 0.79, 0.81).
  At $389~\rm{K}$ the curve with $\lambda=0.79$ describes the data
  over three decades in frequency. The Brillouin 
  peaks have been removed from the data before fitting.}
  \label{ls03}
\end{figure}

\begin{figure}
\epsfxsize=78mm\centerline{\epsffile{\dirfig crn-f03.bb}}
\bigskip
 \caption{
  Characteristic frequency (${\scriptstyle\blacksquare}$) and
  amplitude (${\scriptstyle\square}$) 
  of the susceptibility minimum, extracted from the 
  fits with $\lambda = 0.79$. The rectified plot of $\nu_\sigma^{2a}$
  and $\chi_\sigma^2$ {\em vs.}~$T$ allows to check the MCT predictions
  (\ref{Echisig}) and (\ref{Enusig}). When the analysis is restricted to
  $T<420~$K, linear fits (solid lines) give a consistent
  $T_c$ between 365 and 370~K. Alternatively, a fit to
  $\chi_\sigma^2$  for all but the two lowest temperatures yields
  $T_c\simeq379~$K (dashed line). }
  \label{ls04}
\end{figure}

\begin{figure}
\epsfxsize=78mm\centerline{\epsffile{\dirfig crn-f04.bb}}
\bigskip
 \caption{
  (a) Static structure factor $S(q)$ of 
  CRN (solid line, measured on G41 at the Laboratoire L\'eon Brillouin)
  and of CRN (dotted line, scanned from Ref.~\protect\cite{KaCC95b}),
  both obtained by neutron diffraction at about room temperature.
  (b) Dynamic window of our inelastic scattering experiment on the
  time-of-flight spectrometer IN~5 for an incident neutron
  wavelength $\lambda_i=8.0~{\rm\AA}$. 
  The solid curves show $q(\nu)$ for every second detector. 
  The scattering law $S(q,\nu)$ is obtained by interpolating
  the constant-angle data to constant-$q$ sections. The vertical
  dashed-dotted lines indicate the limits of the $q$ range used in the
  data analysis. }
  \label{ns01}
\end{figure}

\begin{figure}
\epsfxsize=78mm\centerline{\epsffile{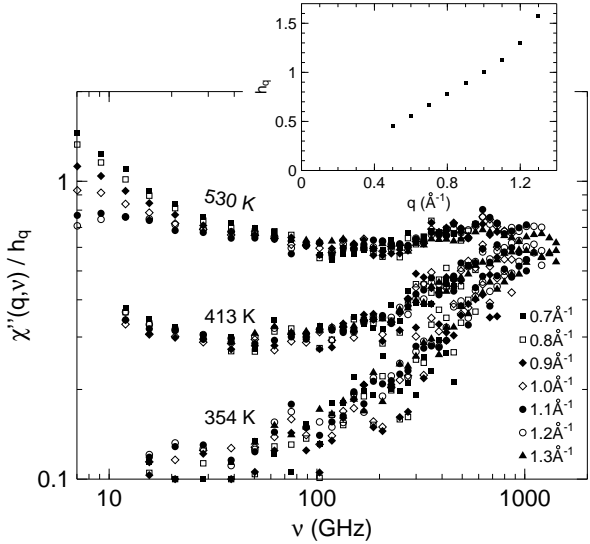}}
\bigskip
 \caption{
  Susceptibilities measured with neutron scattering.
  According to the factorization property of mode-coupling theory the
  dynamics in the fast $\beta$ regime can be described by
  frequency-dependent part and a $q$-dependent factor. Using this
  property the  
  susceptibilities for different $q$ can be collapsed onto a single
  curve by multiplying with a temperature independent factor $h_q$.
  At $530~\rm{K}$ the $q$-dependent $\alpha$ peak moves into the
  experimental window. Apart from the $\alpha$ 
  relaxation all curves can be added to improve the data quality for
  fits to the frequency dependent part. 
  The inset shows $h_q$ as a function of $q$.}
  \label{ns02}
\end{figure}

\begin{figure}
\epsfxsize=78mm\centerline{\epsffile{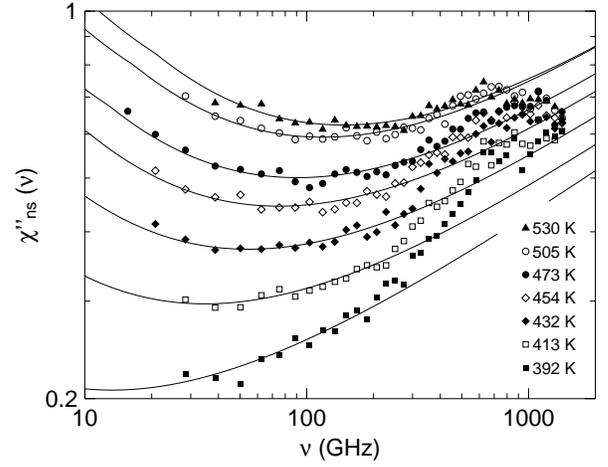}}
\bigskip
 \caption{
  Susceptibility $\chi''_{\rm ns}(\nu) = \left< \chi''(q,\nu)/h_q\right>$
  averaged over $q$ between 0.5 and 1.3$~\rm\AA^{-1}$ (excluding the
  q-dependent $\alpha$ relaxation in the low-frequency,
  high-temperature limit). Solid lines are fits to the MCT asymptote
  with fixed $\lambda=0.79$.}
  \label{ns03}
\end{figure}

\begin{figure}
\epsfxsize=78mm\centerline{\epsffile{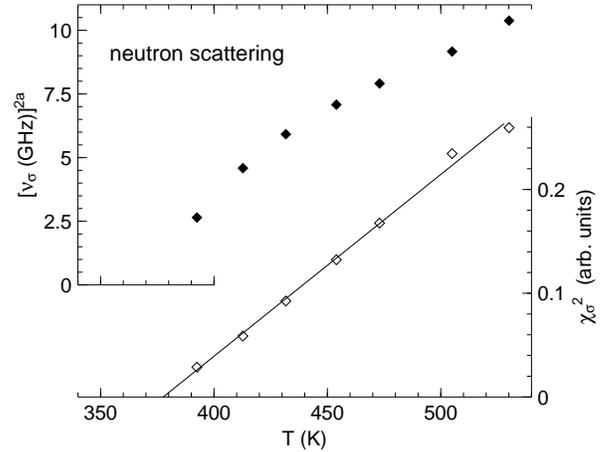}}
\bigskip
 \caption{
  A rectified plot of the $\beta$-relaxation parameters determined by
  neutron scattering does not extrapolate to a
  consistent value for $T_c$. While $\chi_\sigma^2$ suggest an
  extrapolation to $T_c \simeq 380~$K, the $\nu_\sigma^{2a}$ do not
  reach the asymptote (\ref{Enusig}).}
  \label{ns04}
\end{figure}

\begin{figure}
\epsfxsize=78mm\centerline{\epsffile{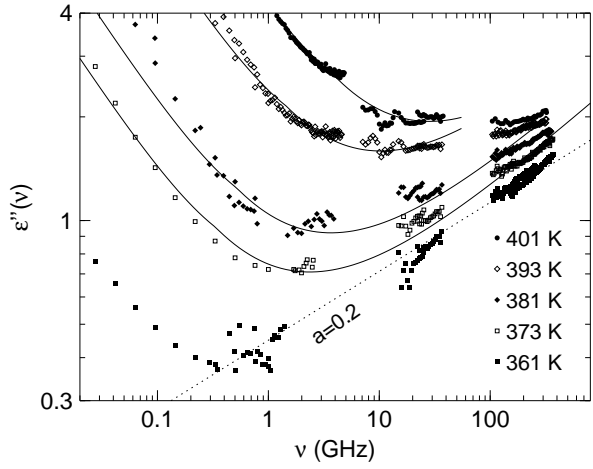}}
\bigskip
 \caption{
  Dielectric loss between 361~K (bottom) and 401~K (top).
  Solid lines show the mode-coupling asymptote 
  with an imposed parameter $\lambda=0.79$.
  These fits work well around the susceptibility minimum,
  but they do not match the exceptionally small slope 
  of the high frequency wing. 
  The dotted line shows the power-law asymptote corresponding to
  $\lambda=0.91$ used in the original publication~\protect\cite{LuPL97}.}
  \label{dk01}
\end{figure}

\begin{figure}
\epsfxsize=78mm\centerline{\epsffile{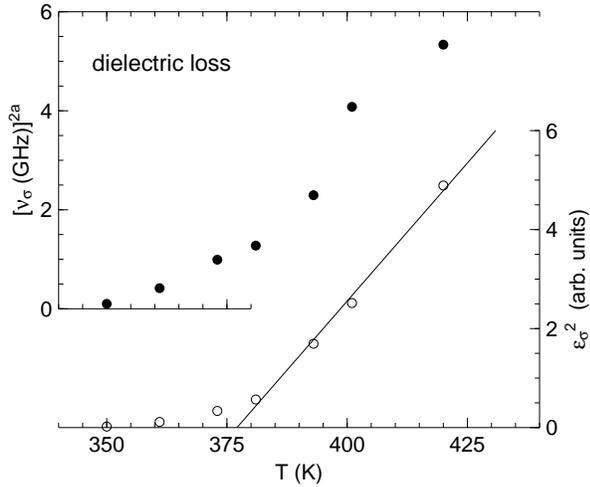}}
\bigskip
 \caption{
  As in Figs.~\ref{ls04} and \ref{ns04}, this plot shows frequency and
  amplitude of fast $\beta$ relaxation, extracted from fits
  with $\lambda=0.79$ and rectified according to the MCT predictions. }
  \label{dk02}
\end{figure}

\begin{figure}
\epsfxsize=78mm\centerline{\epsffile{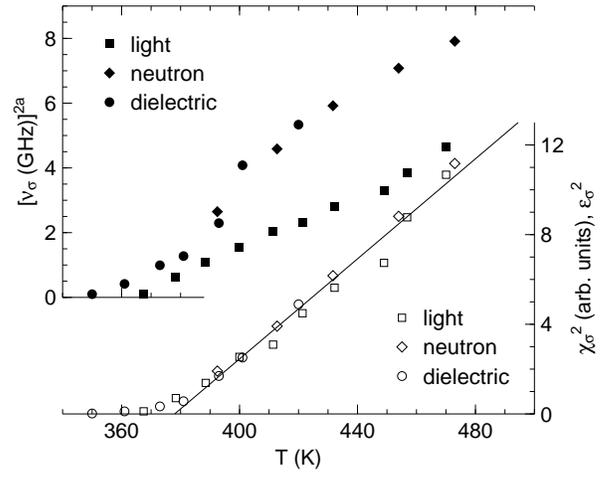}}
\bigskip
 \caption{
  All rectifications combined in one figure. While the amplitudes
  $\chi_\sigma^2$ extrapolate to a consistent $T_c\simeq378~$K for
  all methods, the position $\nu_\sigma$ differ considerably. }
  \label{vg02}
\end{figure}

\begin{figure}
\epsfxsize=78mm\centerline{\epsffile{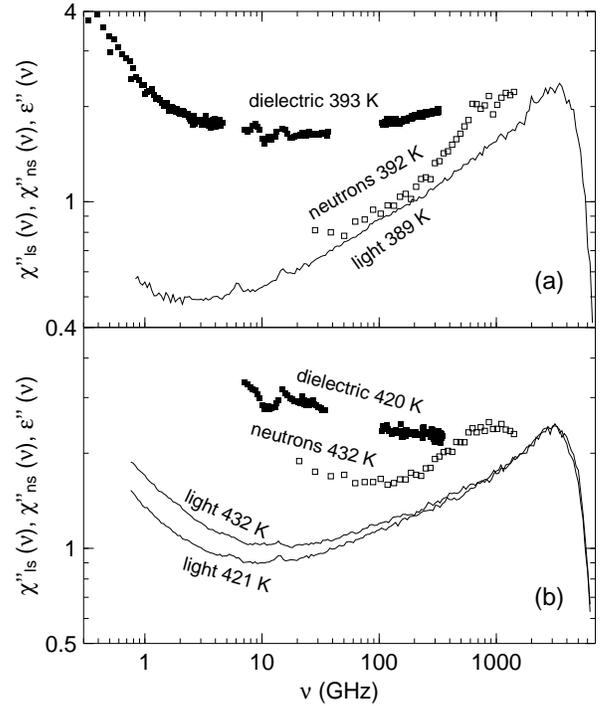}}
\bigskip
 \caption{
  A direct comparison of the three methods on arbitrary susceptibility
  scale:
  (a) all three spectroscopies at about 390~K;
  (b) a pairwise comparison of neutron and light scattering at
  432~K, and of dielectric loss and light scattering at 420~K.
  The minimum positions are grossly different even if 
  each curve for itself seemed describable by the mode-coupling
  asymptote.}
  \label{vg01}
\end{figure}

\end{multicols}
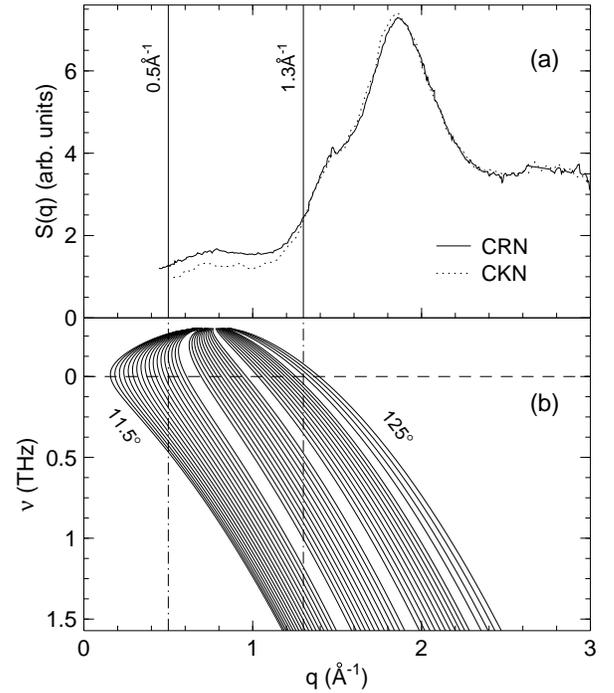
\end{document}